\documentclass{sig-alternate-10pt}
\usepackage{graphicx}
\usepackage{amssymb}
\usepackage{amsmath,url}
\usepackage{multicol,multirow}
\usepackage{epsf,psfrag,pstricks}
\usepackage{rotating,subfigure}
\begin{document}
\title{Where are my followers?\\Understanding the Locality Effect in Twitter}


\numberofauthors{2}
\author{
\small
\alignauthor
Roberto Gonzalez, Ruben Cuevas, Carmen Guerrero\\
\affaddr{Universidad Carlos III de Madrid}\\
{\tt \{rgonza1,rcuevas,guerrero\}@it.uc3m.es}
\alignauthor
Angel Cuevas\\
\affaddr{Institute Telecom, Telecom SudParis}\\
{\tt angel.cuevas\_rumin@it\-sudparis.eu} 
}

\maketitle
\begin{abstract}
Twitter is one of the most used applications in the current Internet with more than 200M accounts created so far. As other large-scale systems Twitter can obtain benefit by exploiting the Locality effect existing among its users. In this paper we perform the first comprehensive study of the Locality effect of Twitter. For this purpose we have collected the geographical location of around 1M Twitter users and 16M of their followers. Our results demonstrate that language and cultural characteristics determine the level of Locality expected for different countries. Those countries with a different language than English such as Brazil typically show a high intra-country Locality whereas those others where English is official or co-official language suffer from an external Locality effect. This is, their users have a larger number of followers in US than within their same country. This is produced by two reasons: first, US is the dominant country in Twitter counting with around half of the users, and second, these countries share a common language and cultural characteristics with US.


\end{abstract}

\section{Introduction}

Twitter \cite{twitter} is a microbloging system created in 2006 by Jack Dorsey and Biz Stone. It has rapidly attracted a large number of users and become one of the most successful platforms for both social interactions and information diffusion. Twitter currently counts with around 200 millions of users and more than 140 millions tweets are uploaded every day to the system. In Twitter a user can post text messages of up to 140 characters named \emph{tweets}. Furthermore, a given user, e.g. Bob,  registered in the system can \emph{follow} any other user in the system, e.g. Alice. We then refer to Bob as an Alice's \emph{follower} and Alice's as a Bob's \emph{friend}. This \emph{friend$\rightarrow$follower} relationship (or link) allows Bob visualizing every \emph{tweet} posted by Alice.

The great success of Twitter has attracted the research community that has recently started to investigate different aspects of Twitter \cite{cha_hamed_twitter,beyond_microbloging,why_we_tweet,bala_twitter,sue_twitter,twitter_group09}. In this paper we study the \emph{Locality} effect in Twitter. This is, we look whether the followers of a given user are geographically concentrated, and if so we identify where. Understanding the Locality phenomenon of large scale systems such as p2p systems \cite{choffnes08:ONO,Cuevas2009:Deepdiving,Picconi09:locality,Xie08:P4P} or Online Social Networks (OSNs) \cite{facebook_locality} is critical in order to improve the system design and the users performance while reducing the infrastructural and operational costs. Furthermore, it can also help on improving the design and performance of the data storage system \cite{josep_sigcomm}.
This paper is, to the best of the authors knowledge, a first step to understand the Locality effect in Twitter.

To conduct our study we have collected a real dataset including the geographical location of around 1M Twitter users (or friends) and more than 16M followers associated to them. Overall, our dataset includes more than 100M of \emph{friend$\rightarrow$follower} links.

We capture the Locality effect with two different metrics: (i) the \emph{link level distance} accounts the distance associated to any \emph{friend$\rightarrow$follower} pair, whereas (ii) the \emph{user level distance} captures a representative metric per user such us the median distance to its followers. Therefore, the main difference is that very popular users (with many followers) weight more at \emph{link level}, while all users have the same influence (median distance) at \emph{user level}.


Using the described metrics we perform a two folk analysis. First, we \emph{look at the forest} as a whole, and second we try to \emph{look at the forest from the trees}. This is, initially we look Twitter as a whole and measure the locality happening at both the \textit{link level} and the \emph{user level}. The obtained results suggest that there is an important intra-country locality effect defined by short-distance links. However, we observe a surprisingly high percentage ($>$ 25\%) of cross-continent relationships, what may imply an \emph{external} Locality phenomenon (this is a user with most of its followers located in a different country). Furthermore, we demonstrate that the level of traditional Locality (i.e. short-distance links) is higher at the \emph{user level} than at the \emph{link level}. This is caused because popular users with a larger number of links are those more likely to experience the described \emph{external} Locality.

In the second part of the paper \emph{we go into the forest} to see if this global trends can be generally applied to every user. For this purpose we perform a country-based analysis. We have selected the country criteria since: first, we observe a high level of intra-country Locality and, second, it allows to accurately group users sharing a language and a culture (which obviously influence the users relationships in Twitter). Specifically, we analyze the 15 countries with a larger number of friends and followers in our dataset. The first result is the predominance of US that is responsible of around half of the friends, followers and links in our dataset. Then, the observed global trends are highly influenced by the Locality properties of US Twitter users. We also analyze for each of the 15 Top countries the locality at the \textit{link level}. For this purpose, we compute the percentage of \emph{friend$\rightarrow$follower} links of the Twitter users of a given country that stay local within the country, go to US and go to a different country than US. We found three different profiles. On the one hand, we have countries experiencing a quite high \emph{intra-country} Locality effect such as Brazil that keep most of the connections local. These countries have typically a different official language than English and a strong and old culture. On the other hand, we found countries that suffer from the \emph{external} Locality phenomenon at the link level. This is, the major portion of its links goes to US. These are those countries where English is official (or co-official) language. Finally, we observe a set of countries that equally share their links among those staying local, those going to US and those going to other countries. Afterwards, we perform the Locality analysis at the user level for 4 countries, US and the most important representative of each of the defined profiles. These are Brazil, UK and France respectively. We confirm that the intra-country Locality grows as follows among these countries Brazil$>$US$>$France$>$UK, which is coherent with the different profiles defined at the \textit{link level}. Specifically, Brazil shows a surprisingly high intra-country Locality, indeed for most of Brazilian users (independently of the popularity) 80 to 90\% of the followers are local. US shows a slightly lower intra-country Locality than Brazil and also at the user level has an important influence on the Locality trends observed for the whole Twitter system. In UK we observe a clear bi-polarity, unpopular users show an important level of intra-country Locality whereas popular users typically experience an external Locality and have most of its followers in the US. Finally, in France we observe a similar bi-polarity with a major bias towards intra-country Locality.

In a nutshell, Twitter locality general trends depict a clear presence of intra-country Locality as well as a non-previously reported external Locality phenomenon. However, these trends cannot generally be
applied since they are mostly influenced by the dominant presence of US in the Twitter demographics (50\% of our dataset). Therefore, studying Locality in Twitter requires a per-country analysis that clearly demonstrates that language and cultural characteristics of a country definitely contribute to its Locality profile. Then, we can find countries such as Brazil with a 90\% of intra-country locality, and some others like Australia where around half of the \emph{friend$\rightarrow$follower} links goes to US while only one quarter are established inside the country.

The rest of the paper is organized as follows: Section \ref{sec:measurements} describes the used measurement methodology as well as the collected dataset. Sections \ref{sec:global} and \ref{sec:country} show the Locality analysis of Twitter at  global and country levels respectively. Finally, Section \ref{sec:rw} presents the related work and Section \ref{sec:conclusion} concludes the paper.


\section{Measurement Methodology and Infrastructure}
\label{sec:measurements}
Our main objective is collecting for a large number of Twitter users its geographical location, the list of its followers and then, the geographical location of them. This information can be obtained from the Twitter REST API \cite{twitter_api}. Specifically, when queried for a given user-id this API provides: $(i)$ the user's profile information including a location-tag introduced by the user, $(ii)$ a list of followers user-ids and $(iii)$ other information such as the number of friends of the user and the number of tweets posted by the user so far.

For our study we have analyzed a random set of 2M users obtained from \cite{sue_twitter}. For each one of these users we have collected the geographical location of the user, the number of friends, posted tweets and followers. Furthermore we have also used the API to find the geographical location of all the followers of each analyzed user. Unfortunately, Twitter limits the number of queries to be performed to 350 per hour per IP address/user-id\footnote{In the past Twitter gifted whitelist accounts which were allowed to perform up to 20K queries per hour. Unfortunately, these whitelist accounts are anymore available.}. Therefore, in order to speed up the  data collection process we developed a master-slave distributed measurement architecture. This architecture counts with 1 master and 20 slaves located in different virtual machines on top of two physical machines. The master indicates to each slave the user-ids to be monitored. Furthermore, each slave has its own IP address and user-id and can then perform 350 queries per hour to the Twitter API. Therefore, by using this distributed measurement architecture we are able to perform up to 7K queries per hour. Finally, the slaves store the collected information into a redundant centralized database.

The collected user's location is the one provided by the user himself in his Twitter profile. Hence, it is not homogeneous and in some cases non-existing or meaningless. Our measurement tool filters those users that do not provide location information or provide a meaningless location. Furthermore we use the Yahoo geolocation API \cite{yahoo_geoapi} in order to homogenize the obtained data. For instance, all those users indicating NY, NYC, New York City, etc are mapped into the same city, i.e. New York City. It is worth to mention that in Appendix \ref{sec:accuracy} we demonstrate that the location-tag provided by the user in its profile accurately defines the geographical location of the user.

We have crawled the Twitter API with the described software from 10-01-2011 until 28-04-2011. The resulting dataset includes (after filtering it) 973K geolocated friends, 16.5M of geolocated followers and more than 100M of \emph{friend$\rightarrow$follower} relationships.


%
\section{Global Locality in Twitter}
\label{sec:global}

\begin{figure}[t]
\centering
\subfigure[Distribution]{\includegraphics[width=1.4in]{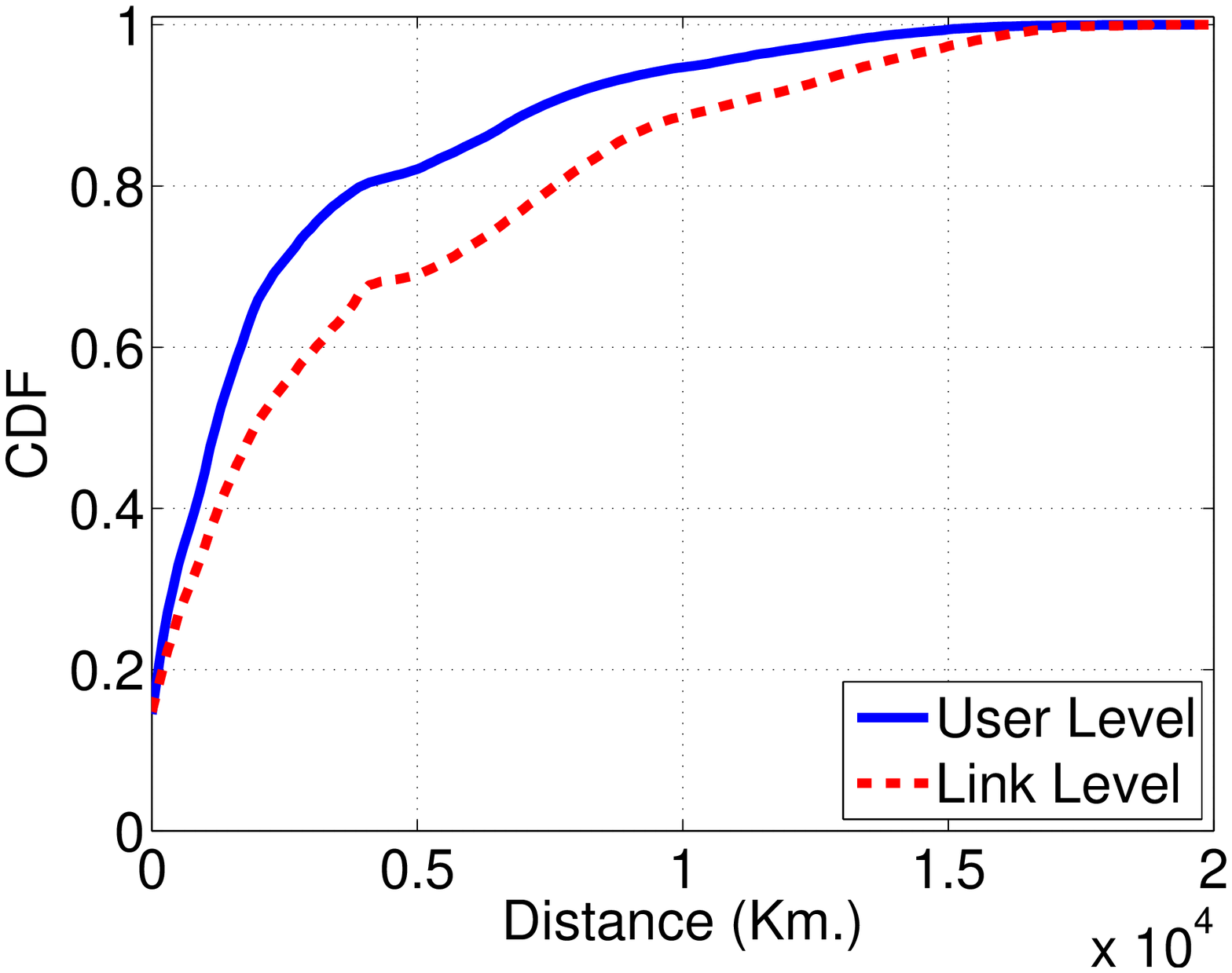} \label{fig:distances}}
\subfigure[Dist. vs Pop.]{\includegraphics[width=1.4in]{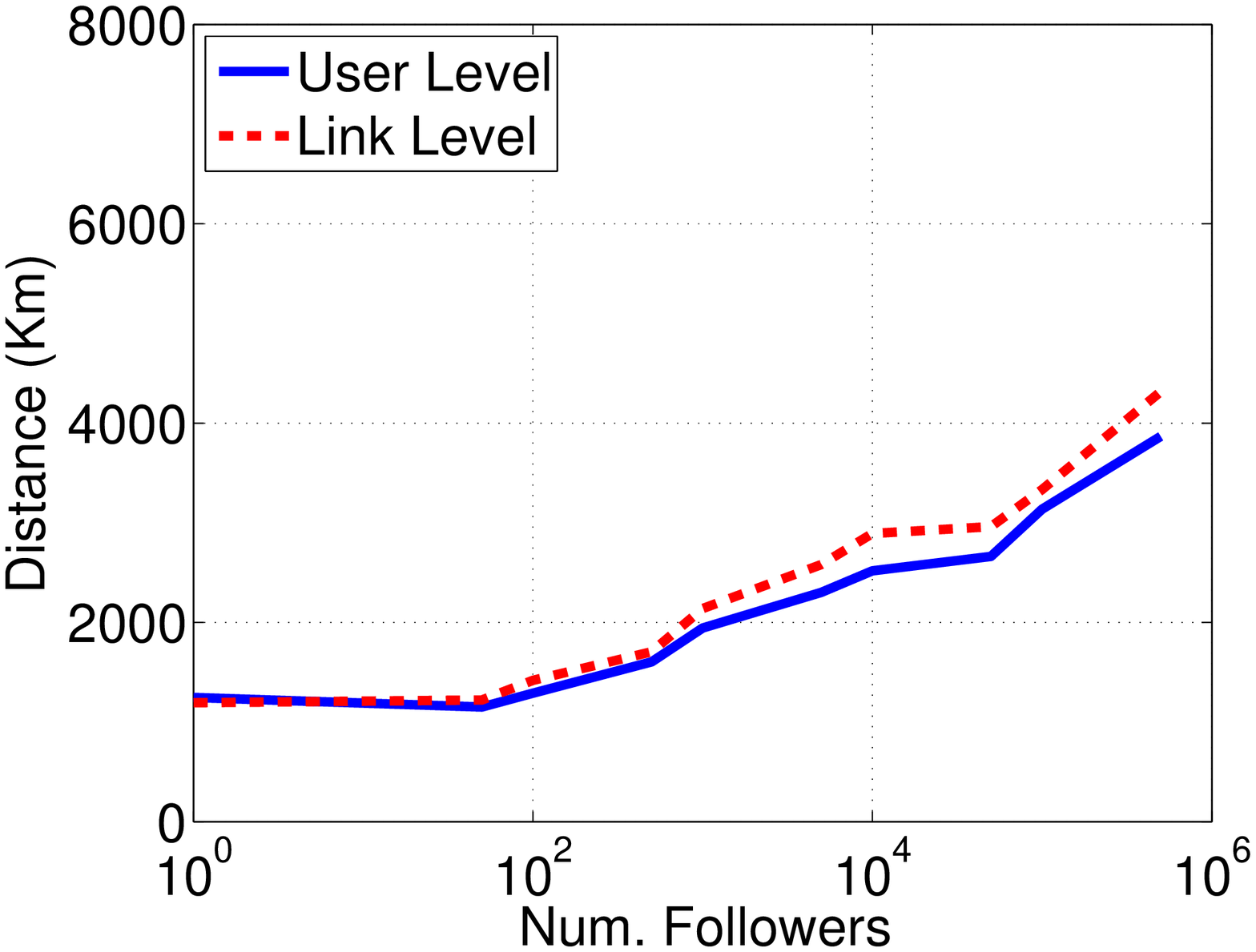} \label{fig:followers}}
\vspace{-0.15in}
\caption{ Distance Distribution and Distance vs popularity for user and link level Locality}
\label{fig:distances_and_distribution}
\vspace{-0.15in}
\end{figure}



In this section we quantify the level of Locality of the \emph{friends$\rightarrow$followers} graph in Twitter. This is, we aim to answer the following question: \emph{Are followers typically located close to its friends?}. For this purpose we use the two following metrics:

\noindent \emph{-link level distance}: the geographical distance for each individual \emph{friend$\rightarrow$follower} link in our dataset. 

\noindent \emph{-user level distance}: the median geographical distance between a friend and its followers population. 

Figure \ref{fig:distances} represents the CDF for both metrics. If we focus first on the \emph{link level distance}, we observe that 35\% of the links have an associated distance lower than 1000 km. This represents  intra-country communications for the most representative countries in our dataset (See Tab \ref{tab:countries}). Furthermore, we observe that 67\% of the links are in a range of 4000 km, which means intra-country communications for big countries such as US or Brazil and intra-continent relationships for western Europe. However, there is still around 25\% of long-distance links over 6500 km that represent cross-continent links. Therefore, we can conclude that Twitter is not a very highly localized system. It must be noted that the \emph{link level distance} analyzes individual links and does not capture well the Locality at the user level, since popular users with millions of followers have a higher impact in the presented distribution than those unpopular users. The \emph{user level distance} instead, avoids this effect. We observe that the distribution at the user level is more skewed than the previous one. Specifically, 80\% of the users have a typical distance to its followers $\leq$ 4000 km (i.e. intra-country or intra-continent links). Hence, the user level depicts a higher intra-country locality than the link level.  This suggests that popular users (i.e. those with a larger number of followers) are responsible for most of the long-distance links and has a typical distance to its followers larger than those unpopular users. In order to confirm this hypothesis we group the users by its popularity\footnote{We group the users in the following popularity buckets as function of the number of followers: [1-50],[51-100],[100-500],[501-1000],[1001-5000],[5001-10000], [10001-50000], [50001-100000], [100001-500000] and a last bucket including all those users having $>$ 500K followers.} (i.e. number of followers) and for each group we calculate the median for user and link level distances. The results are depicted by Figure \ref{fig:followers}. The figure validates the previous hypothesis, since we observe that the more popular a user is, the larger is also the distance to its followers population. 

\begin{table}[t]
\tiny
\begin{center}
\begin{tabular}{l||l|l|}
Country &  Friends   & Followers      \\
	     &  (num / \%)   & (num / \%)  \\\hline\hline
US	& 528K / 54.24\%  & 7.37M / 44.59\%  \\
UK	& 70.6K / 7.27\%    & 987K / 5.98\%  \\
BR	& 61.7K / 6.34\%    & 1.81M / 10.94\% \\
CA	& 39.4K / 4.05\%  & 565K / 3.42\%\\
DE	& 21.7K / 2.23\%    & 331K / 2.00\%  \\
AU	& 20.3K / 2.09\%    & 232K / 1.40\%   \\
IN	& 18.8K / 1.93\% & 442K / 2.67\%    \\
NL	& 14.9K / 1.53\%    & 334K /	2.02\%   \\
ID	& 12.1K / 1.24\%    & 862K / 5.22\% \\
FR	& 10.8K / 1.11\%    & 232K /	1.41\%  \\
ES	& 8.7K / 0.89\%     & 277K /	1.68\%  \\
IT	& 7.1K / 0.73\%     & 159K /	0.96\%  \\
JP	& 6.9K / 0.71\%     & 192K /	1.16\%   \\
IE	& 6.5K / 0.67\%     & 95.4K /	0.58\%   \\
MX	& 5.5K / 0.56\%    & 234K /	1.41\%  \\\hline\hline
TOP 15& 833K / 85.60\% & 13.37M / 85.44\%  \\
ALL & 973K / 100\%     & 16.53M	/ 100\% \\\hline\hline

\end{tabular}
\end{center}
\vspace{-0.15in}
\caption{Number of friends and followers of the Top 15 countries in our dataset} 
\label{tab:countries} 
\end{table}

\begin{figure}[t]
\centering
\includegraphics[width=2in]{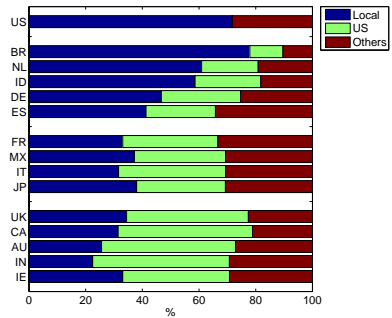}
\vspace{-0.15in}
\caption{Geographical distributions of the followers for the Top 15 countries (percentage of followers within the country, in US and in other countries)}
\vspace{-0.15in}
\label{fig:countries_interactions}
\end{figure}

In summary, we have demonstrated that Twitter is not a highly localized system at the link level since there is an important portion of long-distance relationships whereas the localization is more marked at the user level. Furthermore, we have seen that popularity clearly impacts the Locality level of the users. 
However, this global analysis is clearly influenced by the dominance of US that represents 50\% of the friends, followers and links in our dataset (See Table \ref{tab:countries}). Therefore, 
in the rest of the paper we will deepen and broad the study by analyzing geo-political, cultural and language aspects in order to answer the following questions: \emph{Are the reported global observations valid for every country?} \emph{What are the causes of the observed distribution of intra-country, intra-continent and cross-continent relationships?}.

\begin{figure*}[t]
\centering
\subfigure[US]{\includegraphics[width=1.4in]{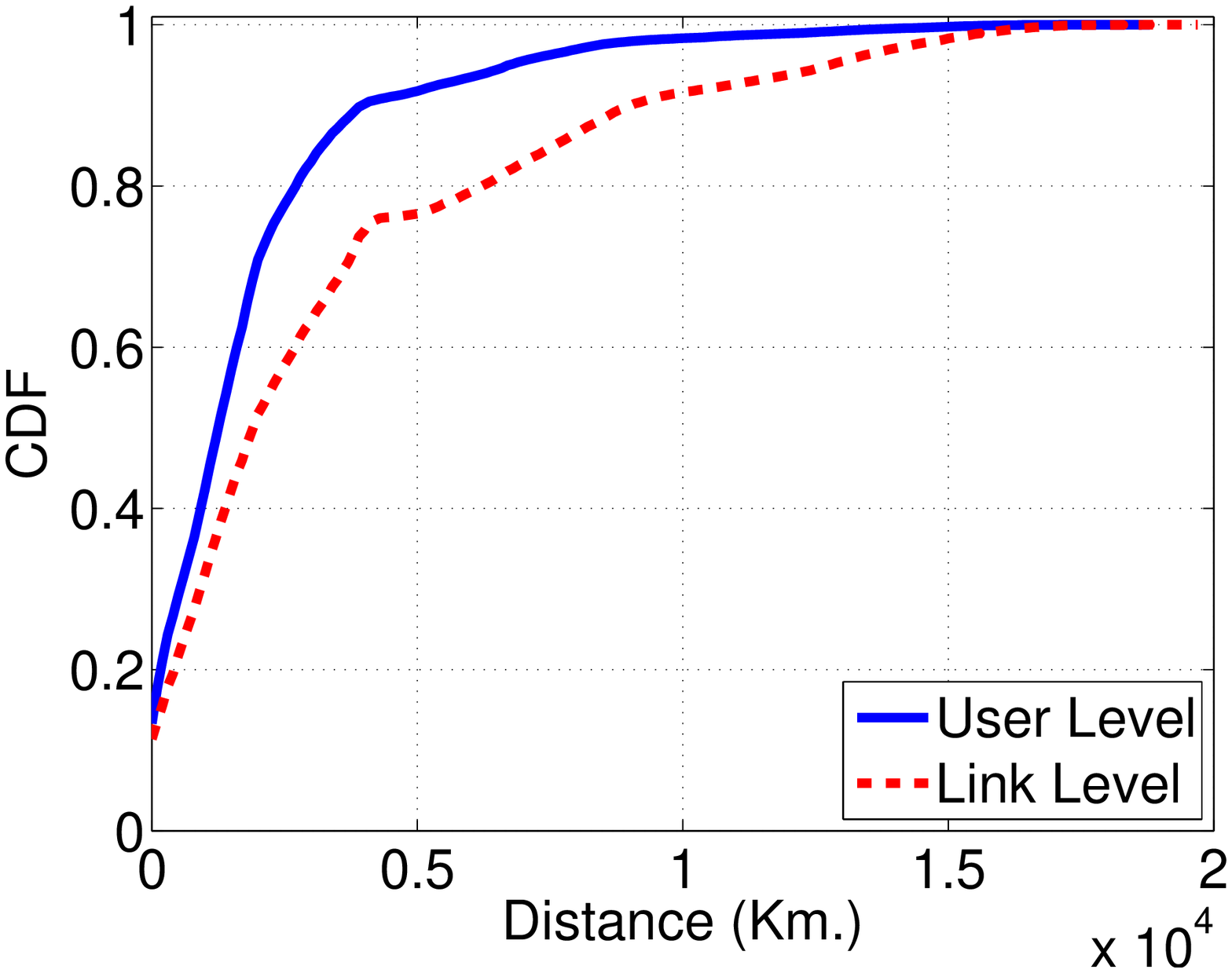} \label{fig:dist_all_US}}
\subfigure[UK]{\includegraphics[width=1.4in]{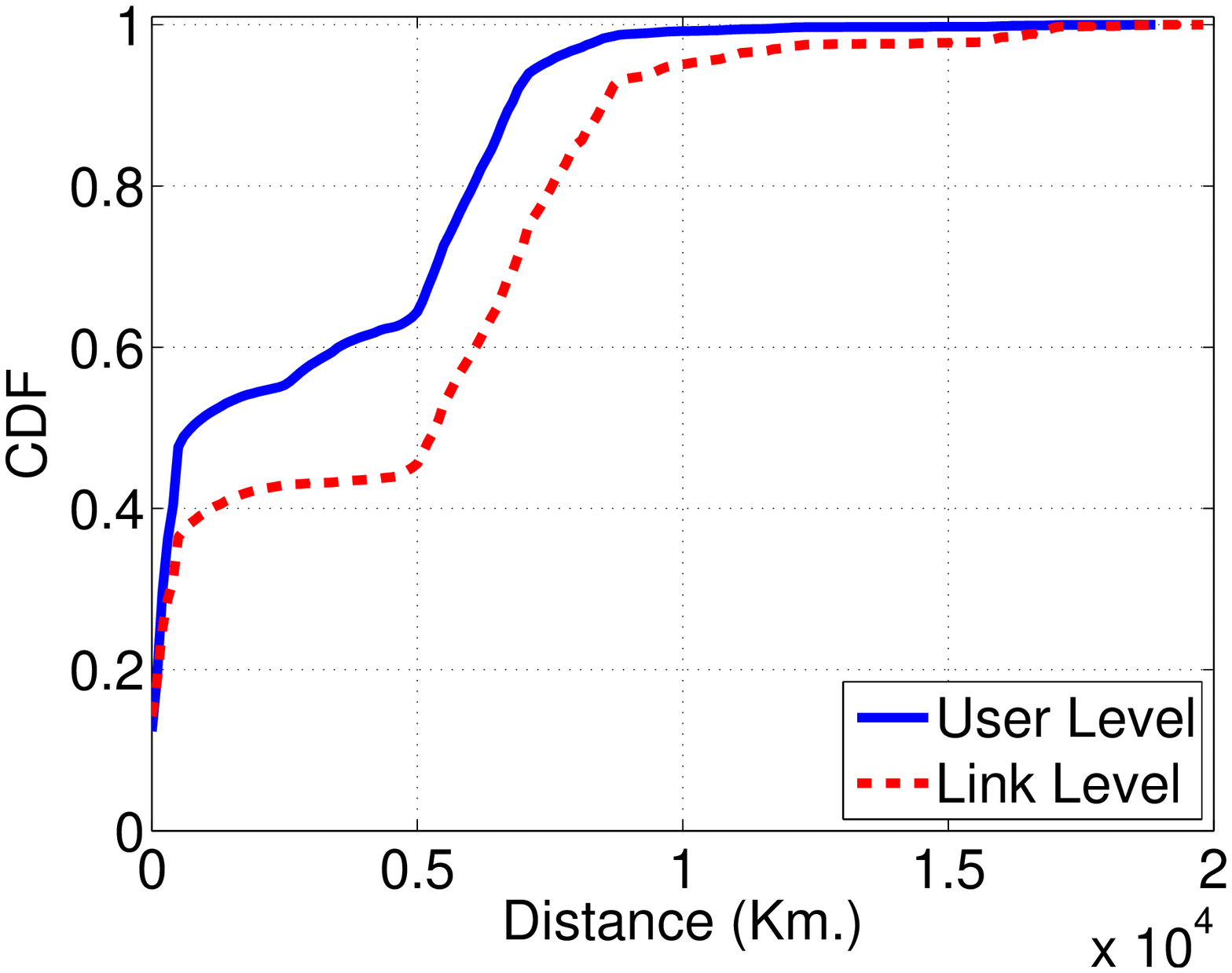} \label{fig:dist_all_GB}}
\subfigure[FR]{\includegraphics[width=1.4in]{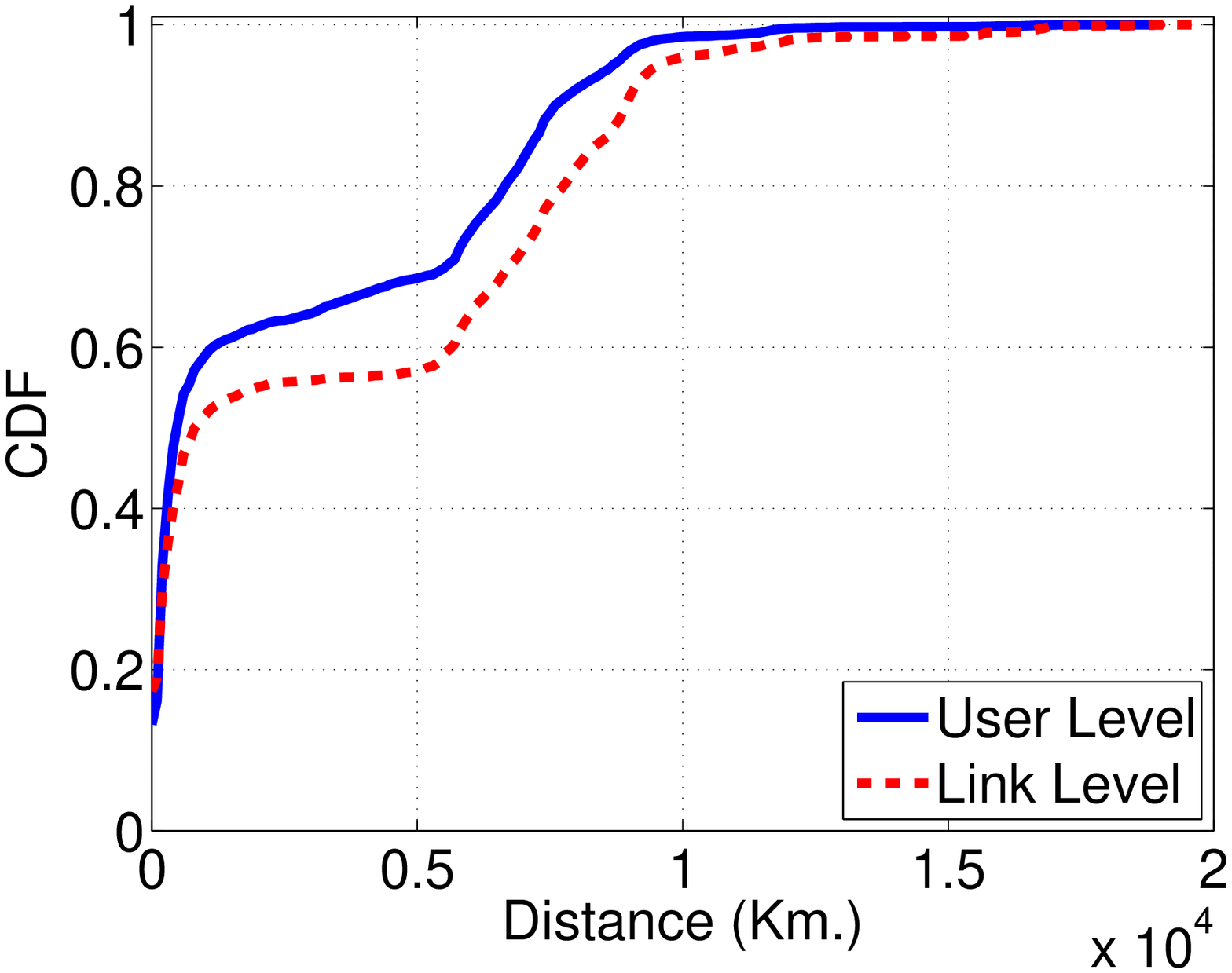} \label{fig:dist_all_FR}}
\subfigure[BR]{\includegraphics[width=1.4in]{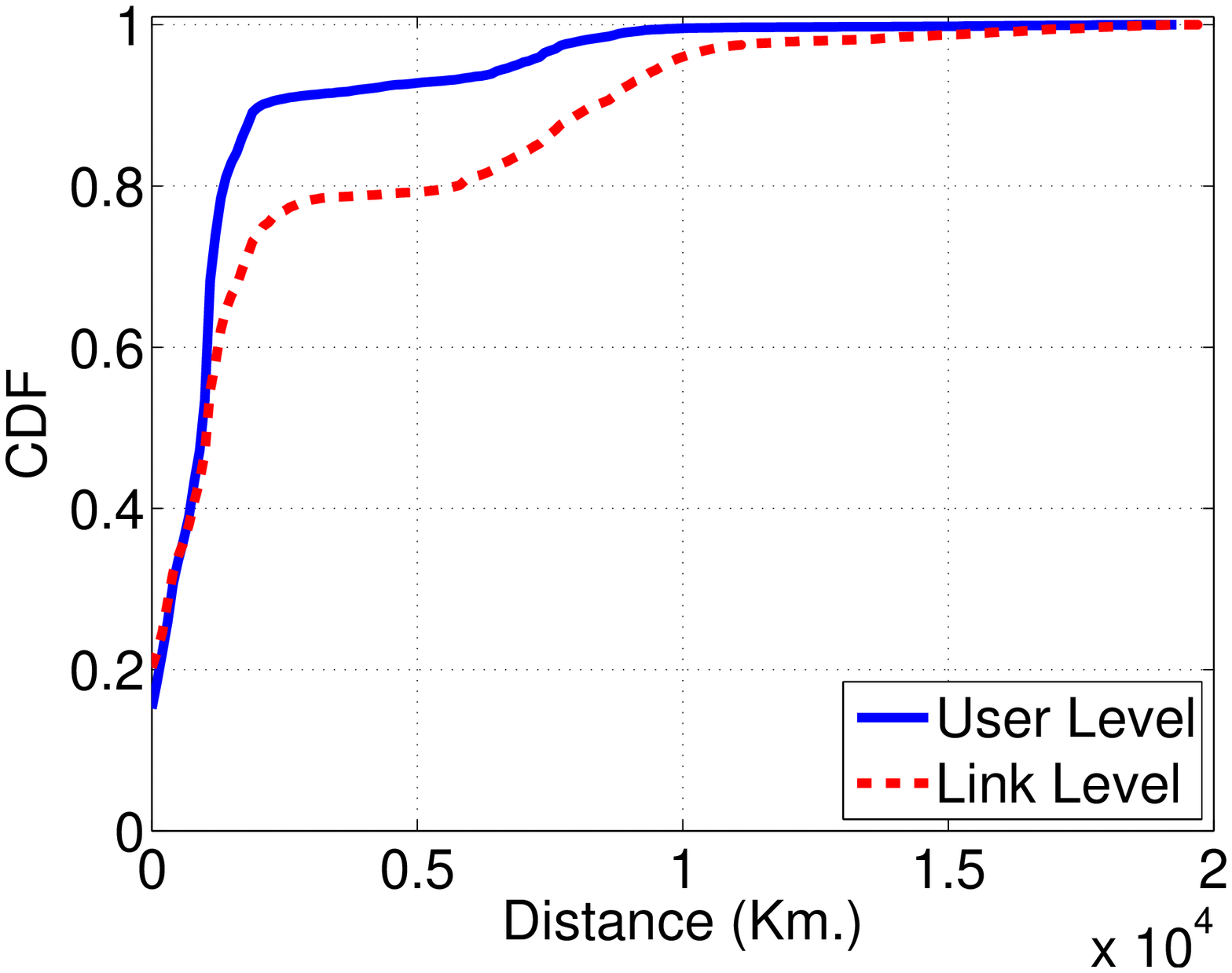} \label{fig:dist_all_BR}}

\vspace{-0.15in}
\caption{Distance Distribution for user and link level Locality: US, UK, France and Brazil}
\label{fig:distances_countries}
\end{figure*}

\begin{figure*}[t]
\centering
\subfigure[US]{\includegraphics[width=1.4in]{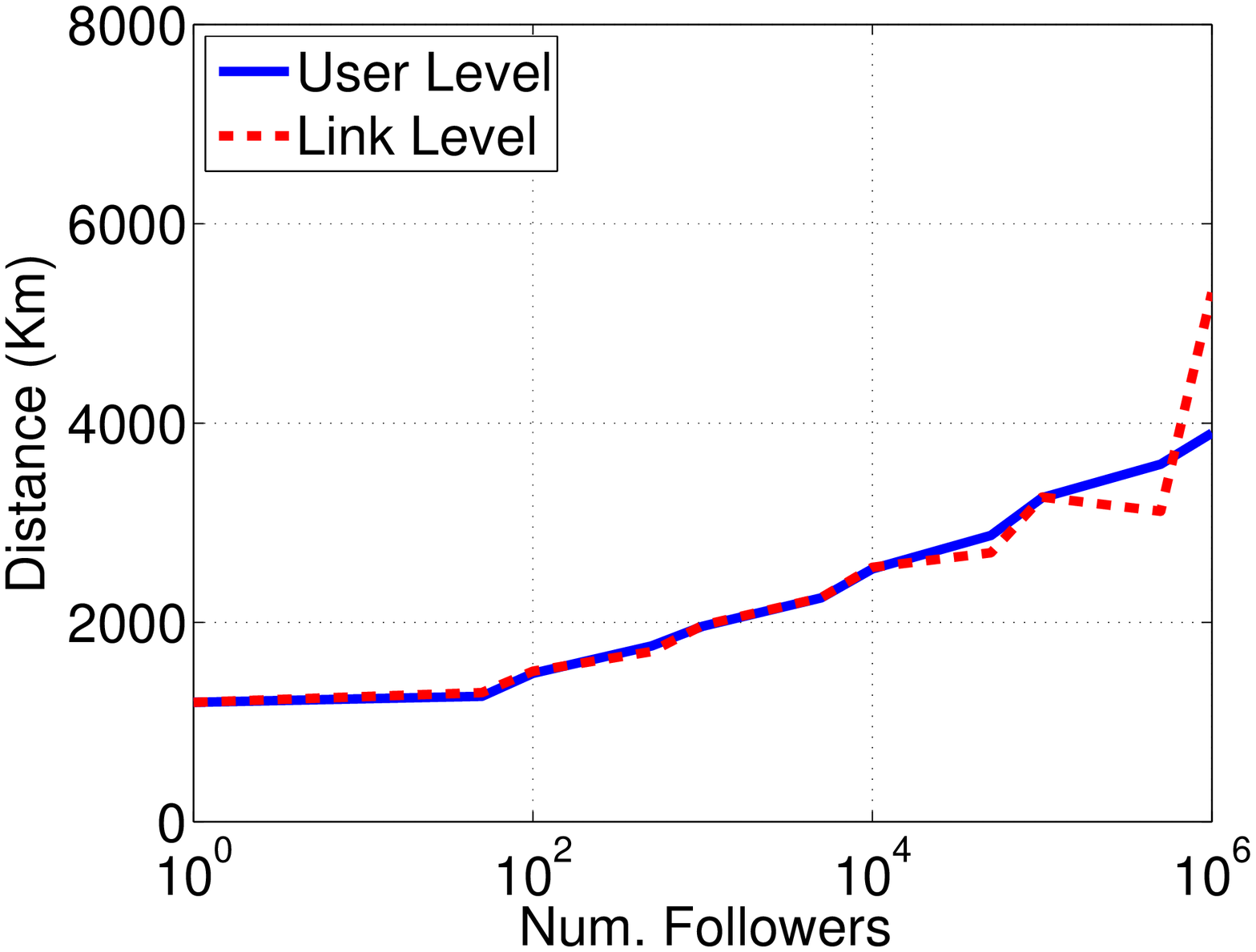} \label{fig:followers_all_US}}
\subfigure[UK]{\includegraphics[width=1.4in]{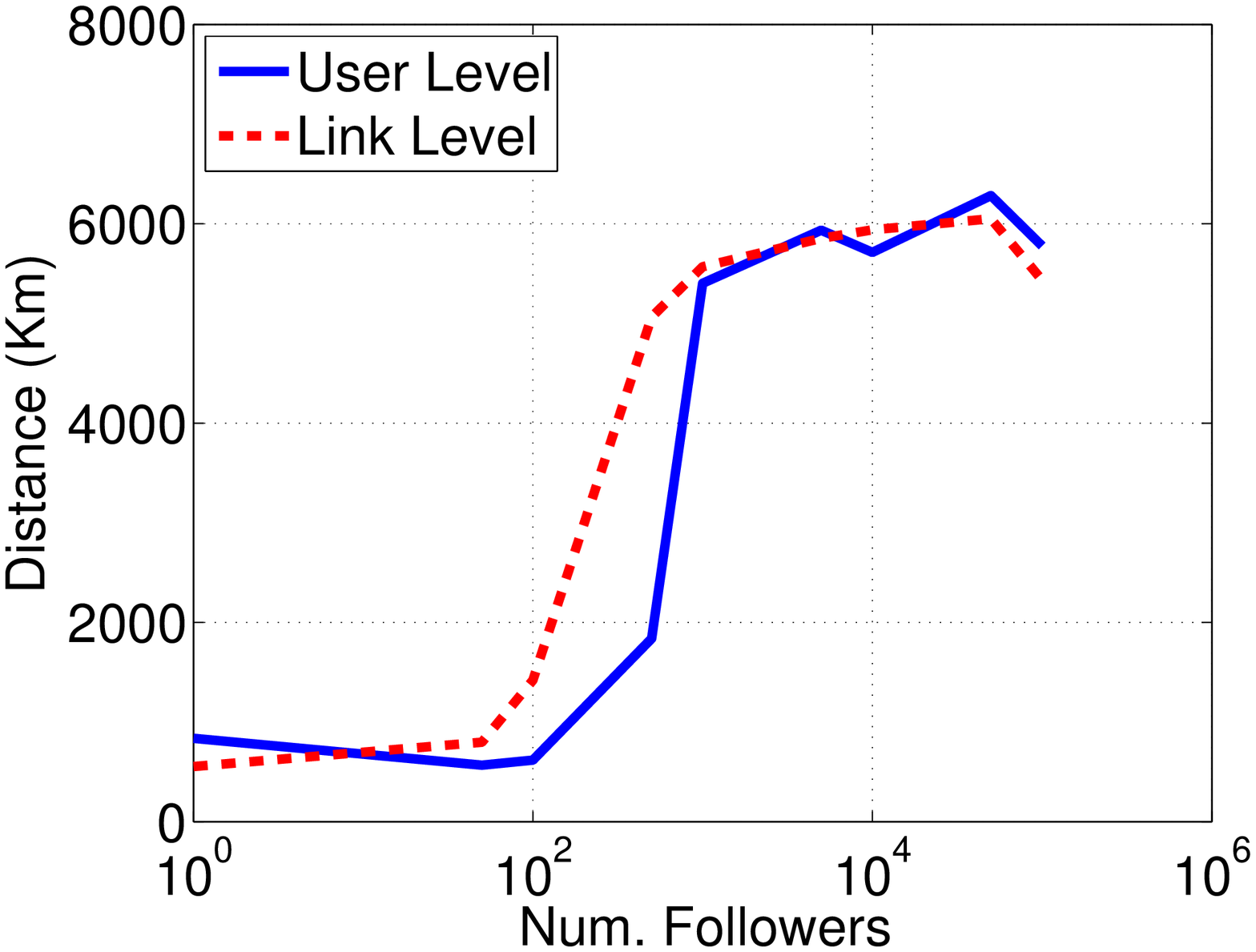} \label{fig:followers_all_GB}}
\subfigure[FR]{\includegraphics[width=1.4in]{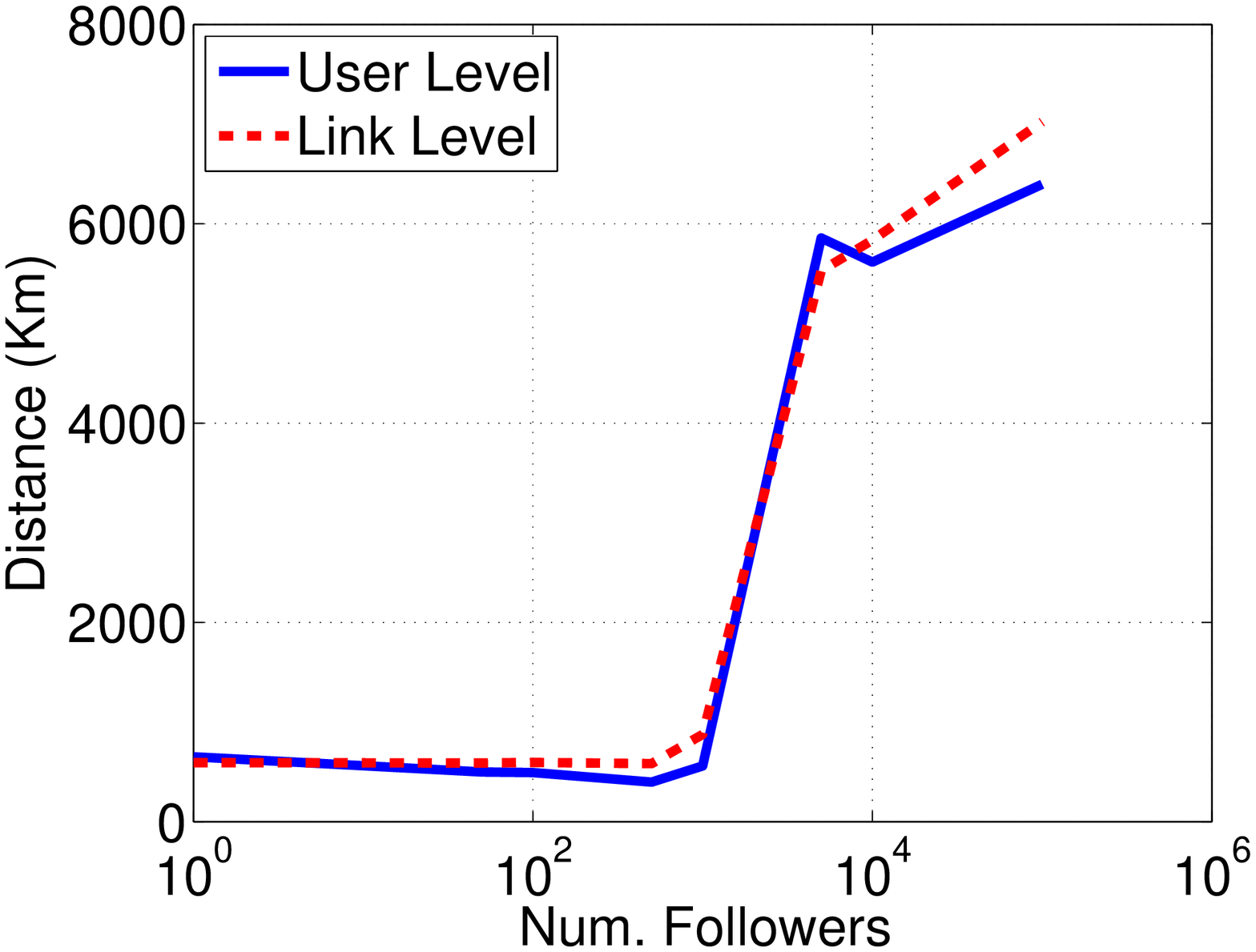} \label{fig:followers_all_MX}}
\subfigure[BR]{\includegraphics[width=1.4in]{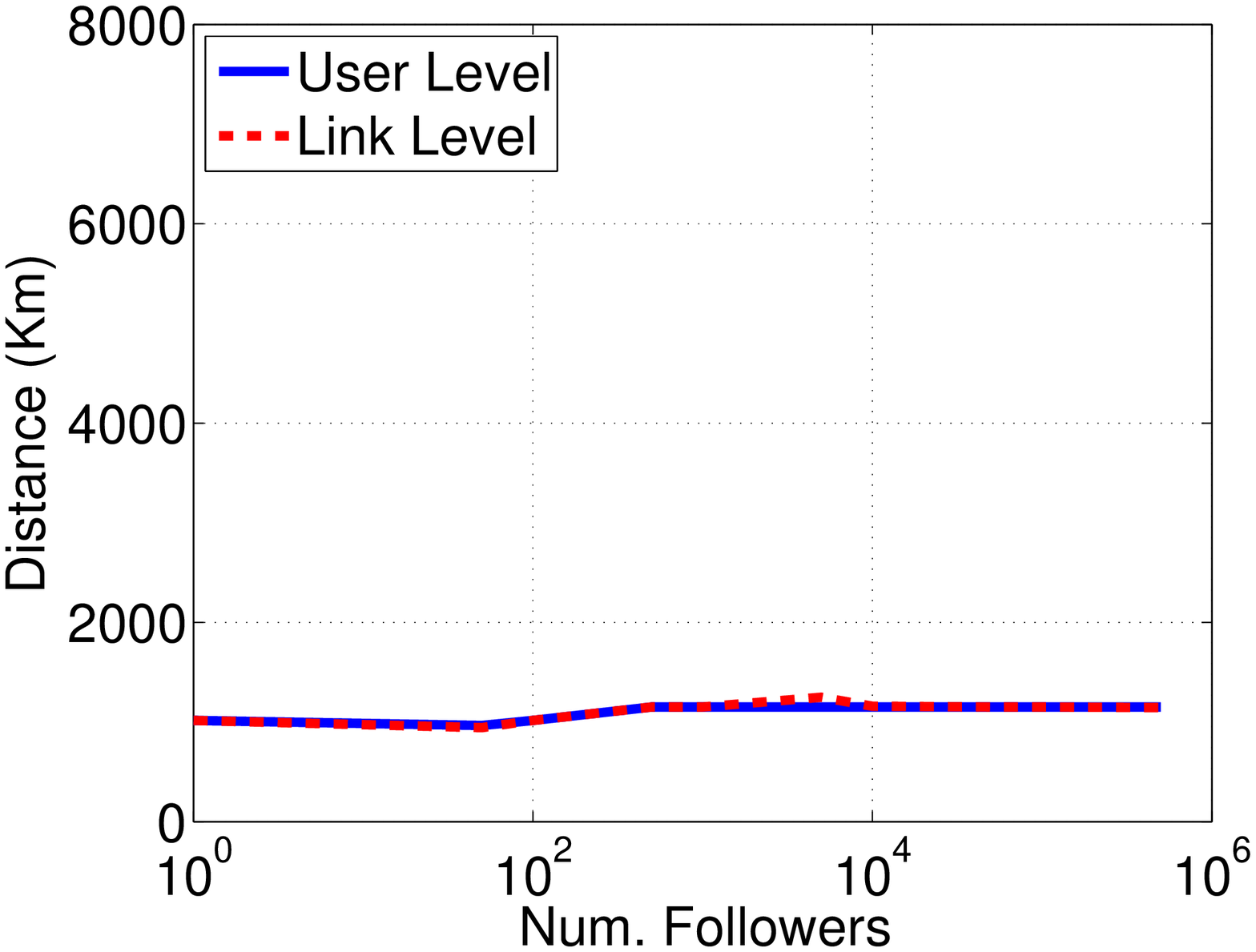} \label{fig:followers_all_BR}}


\vspace{-0.15in}
\caption{Distance vs Popularity for user and link level Locality: US, UK, France and Brazil}
\label{fig:followers_countries}
\vspace{-0.10in}
\end{figure*}

\begin{figure*}[t]
\centering
\subfigure[UK]{\includegraphics[width=1.4in]{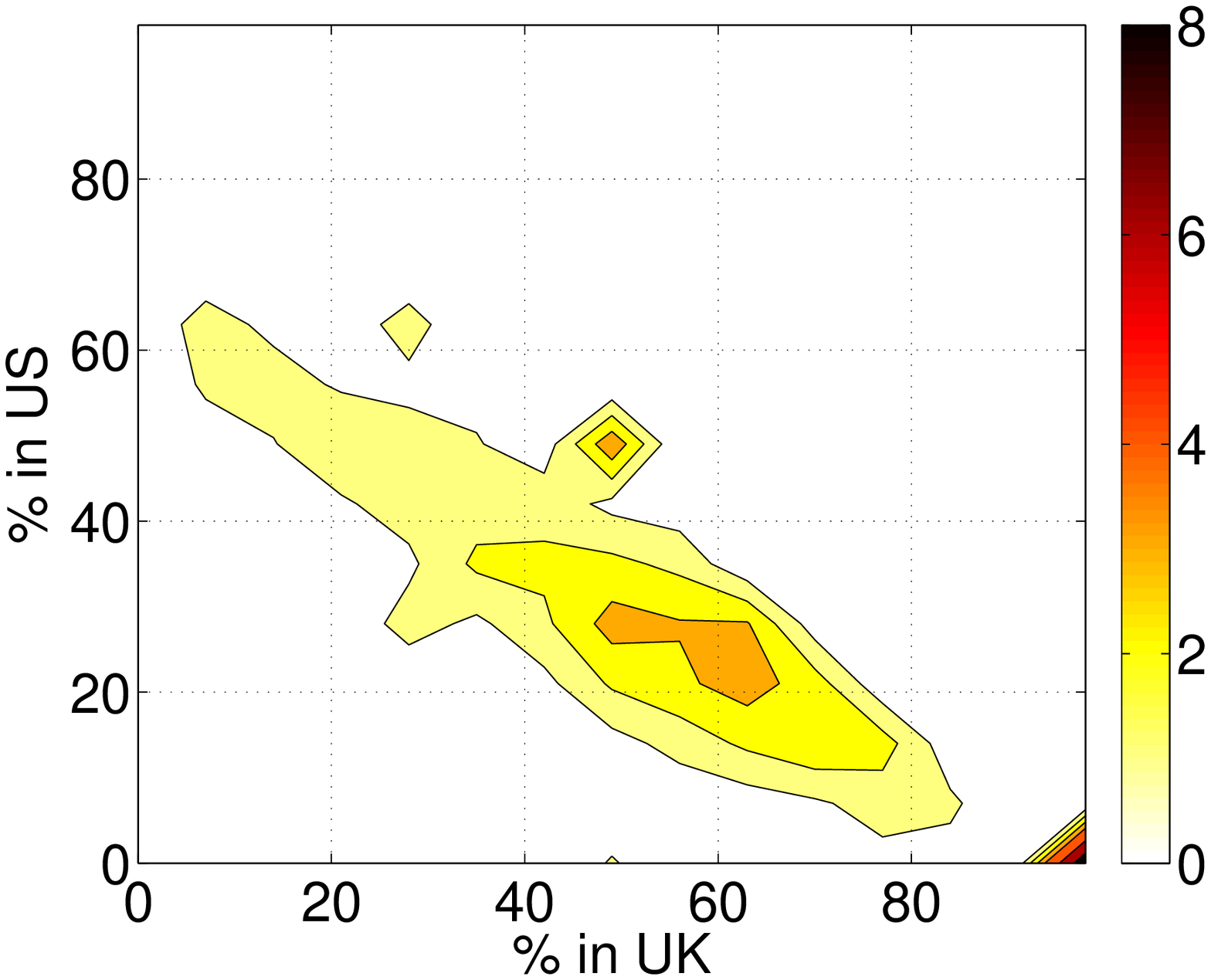} \label{fig:countries_GB}}
\subfigure[FR]{\includegraphics[width=1.4in]{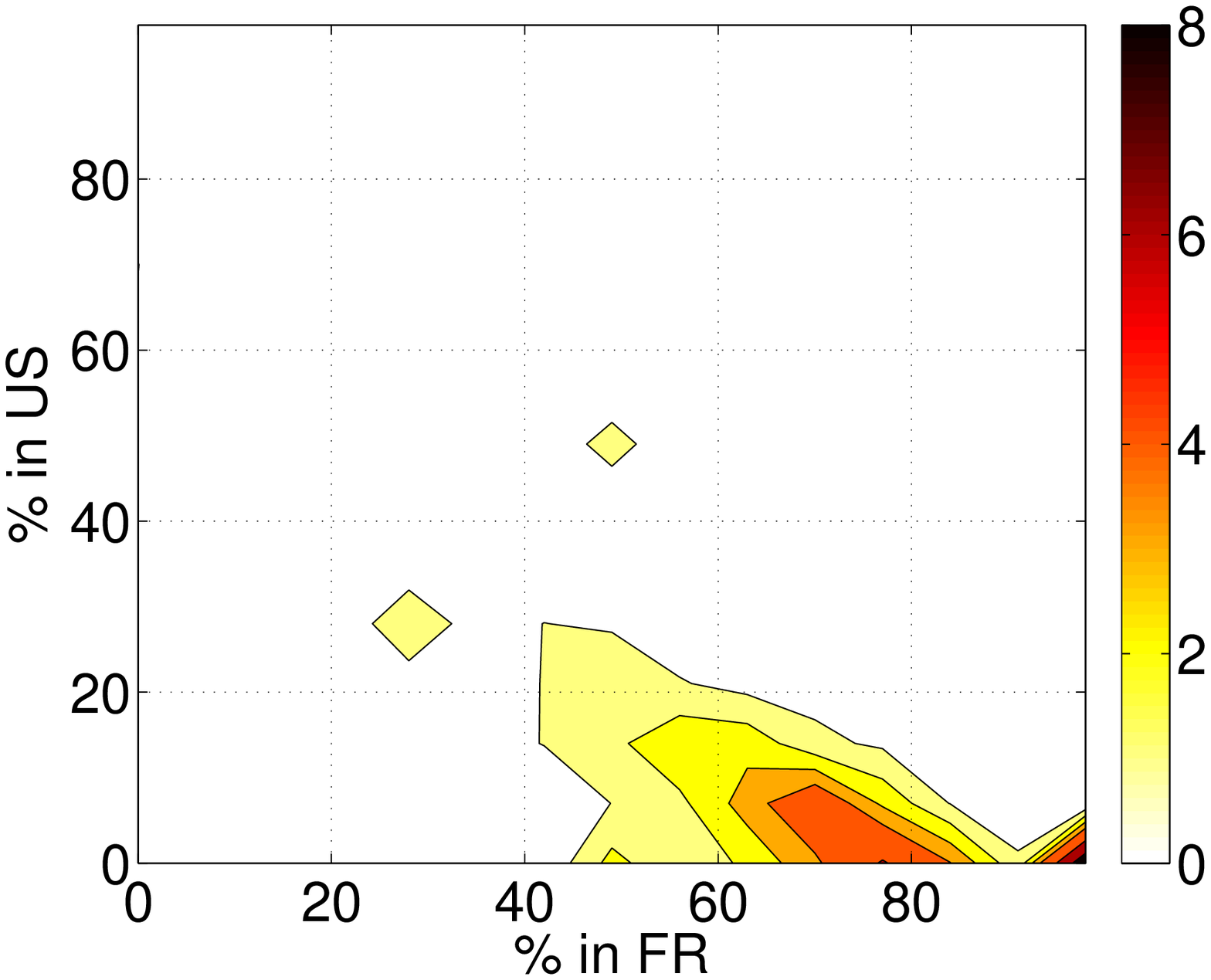} \label{fig:countries_FR}}
\subfigure[BR]{\includegraphics[width=1.4in]{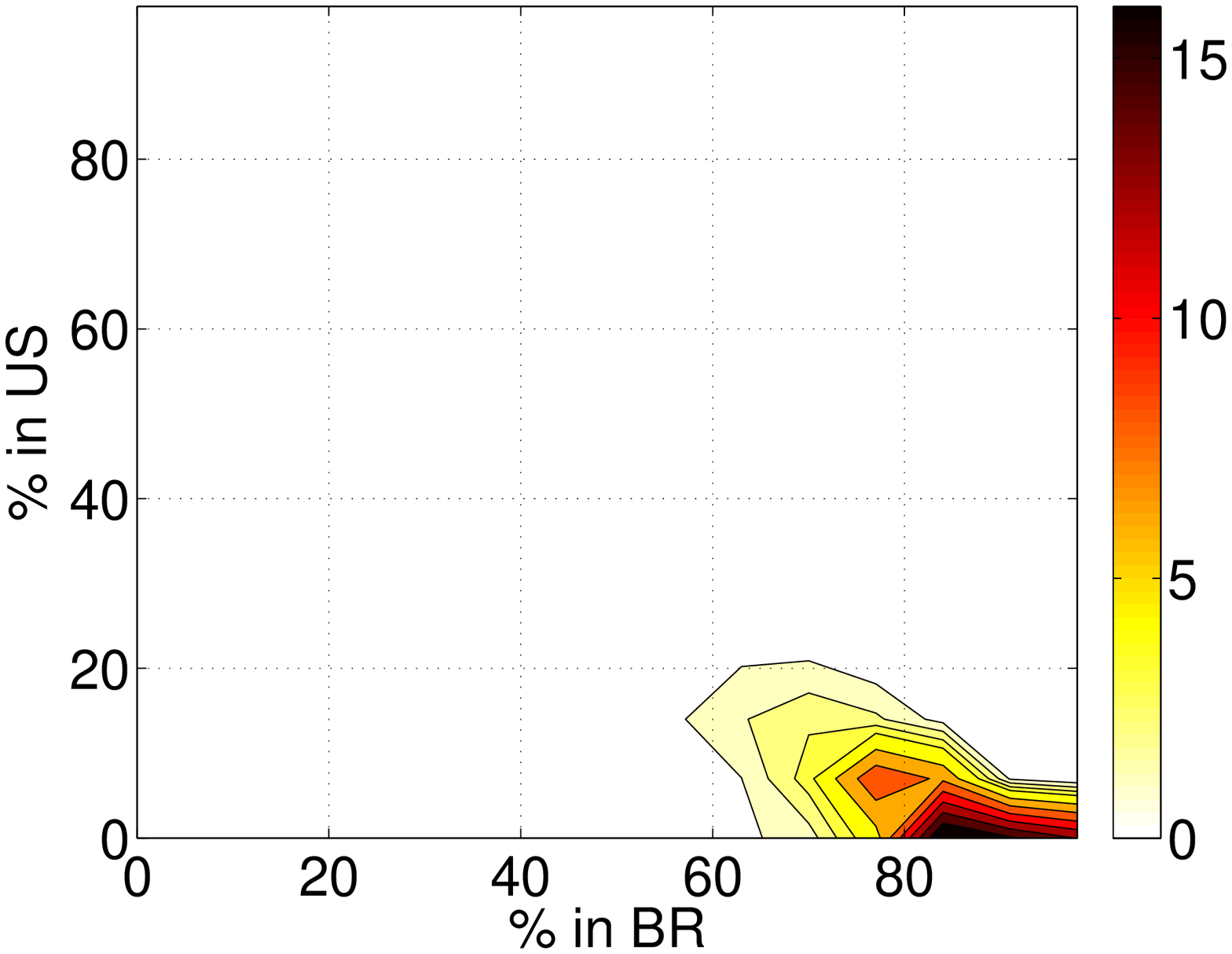} \label{fig:countries_BR}}

\subfigure[US]{\includegraphics[width=1.4in]{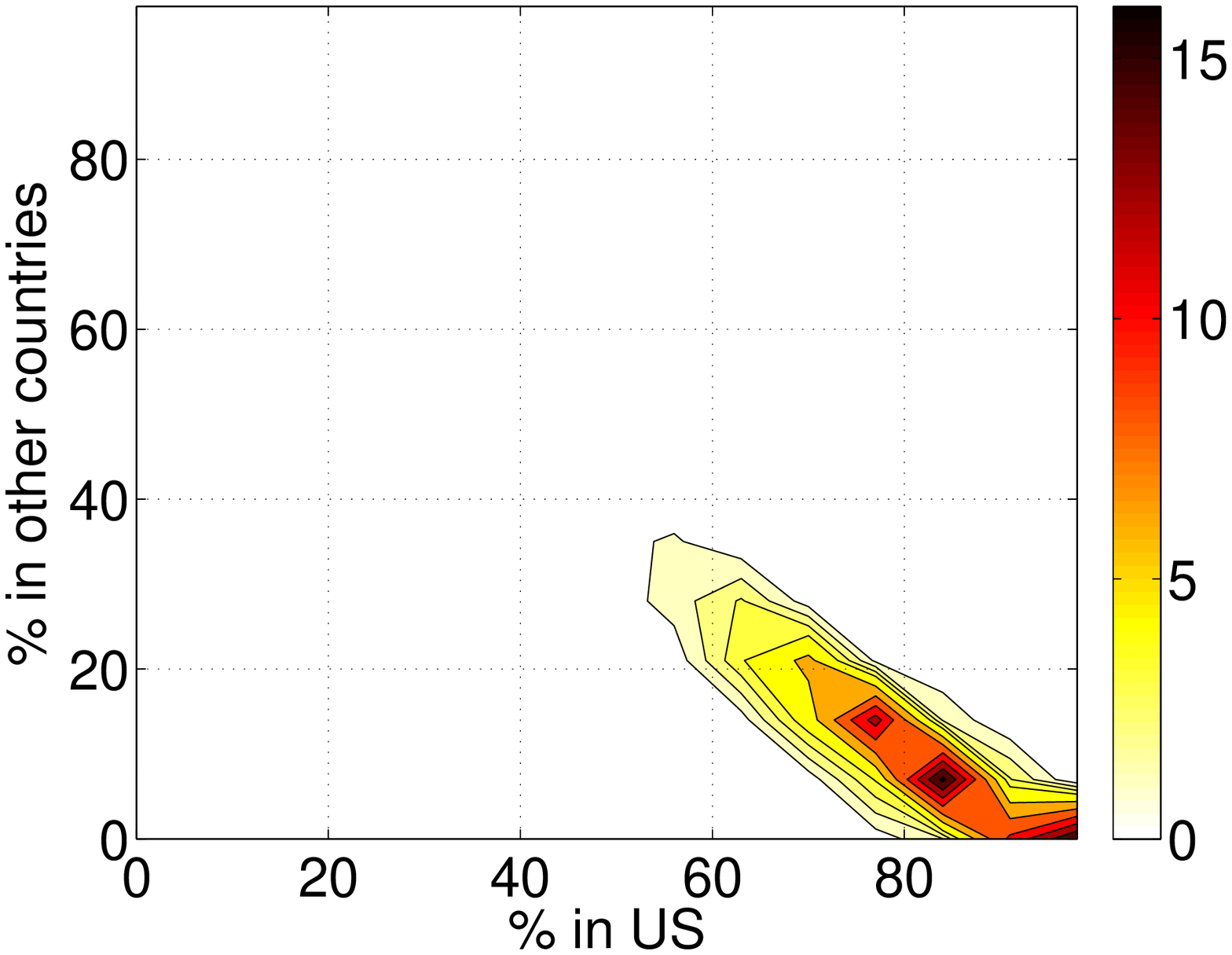} \label{fig:countries_US}}
\subfigure[UK]{\includegraphics[width=1.4in]{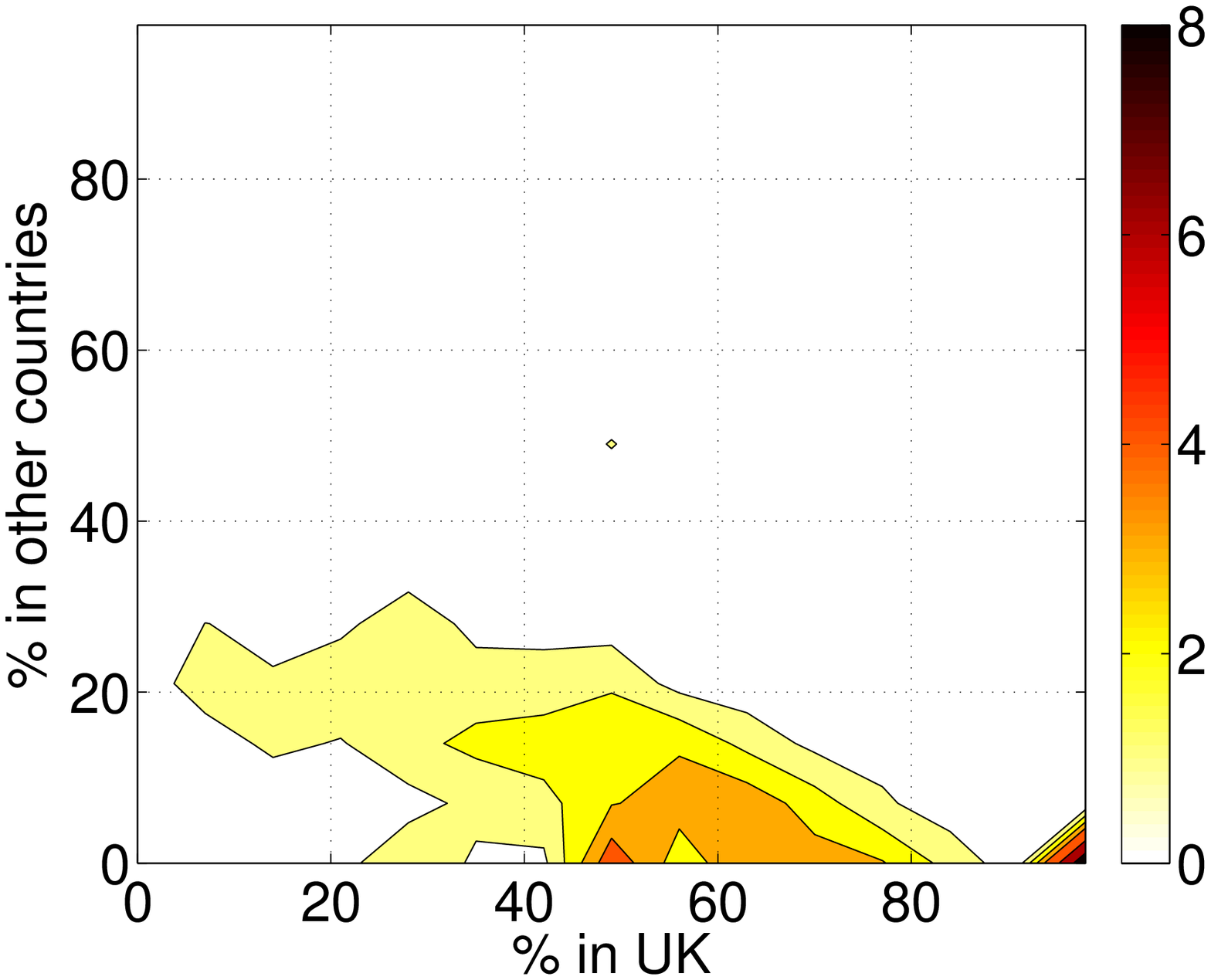} \label{fig:countries_GB_o}}
\subfigure[FR]{\includegraphics[width=1.4in]{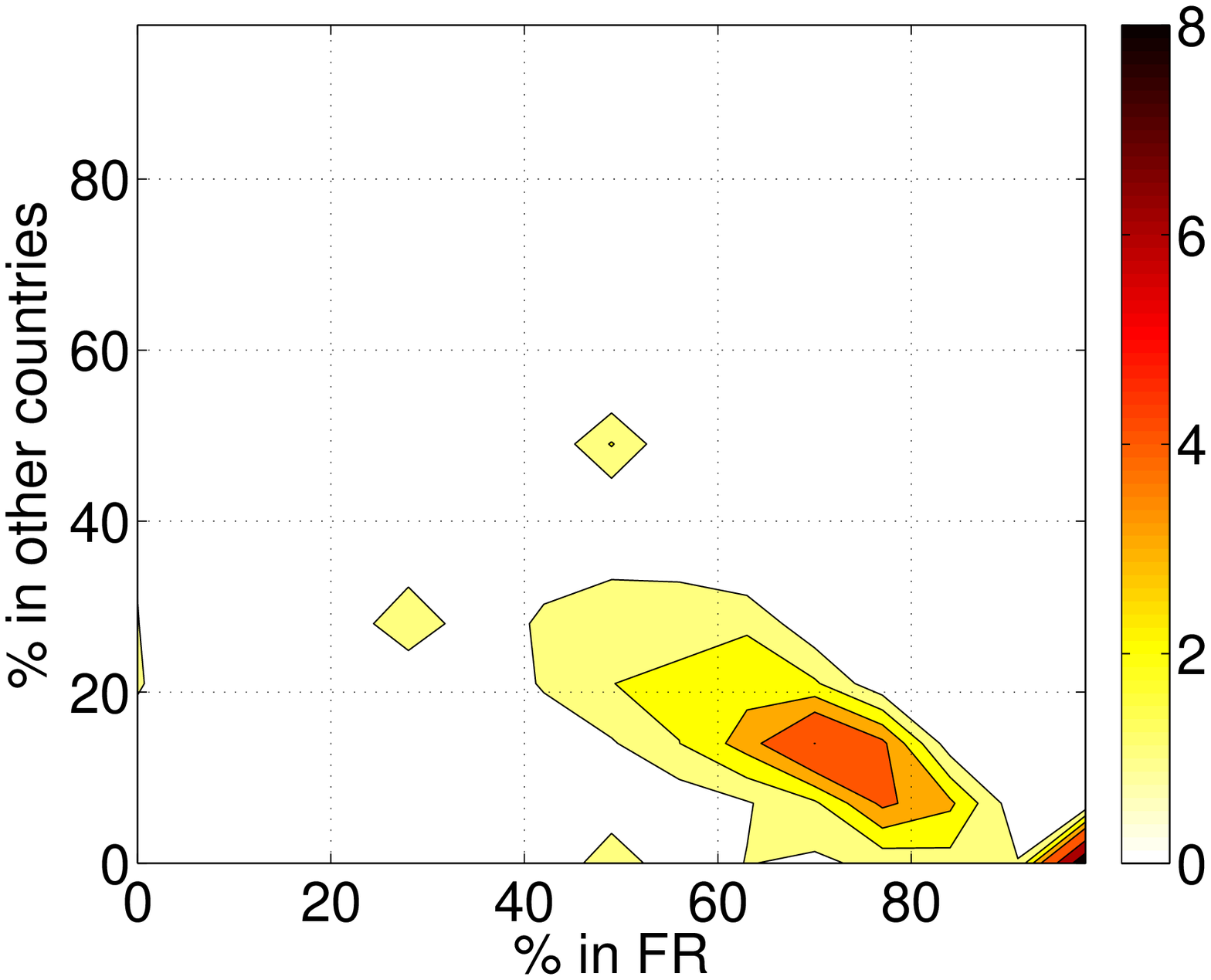} \label{fig:countries_FR_o}}
\subfigure[BR]{\includegraphics[width=1.4in]{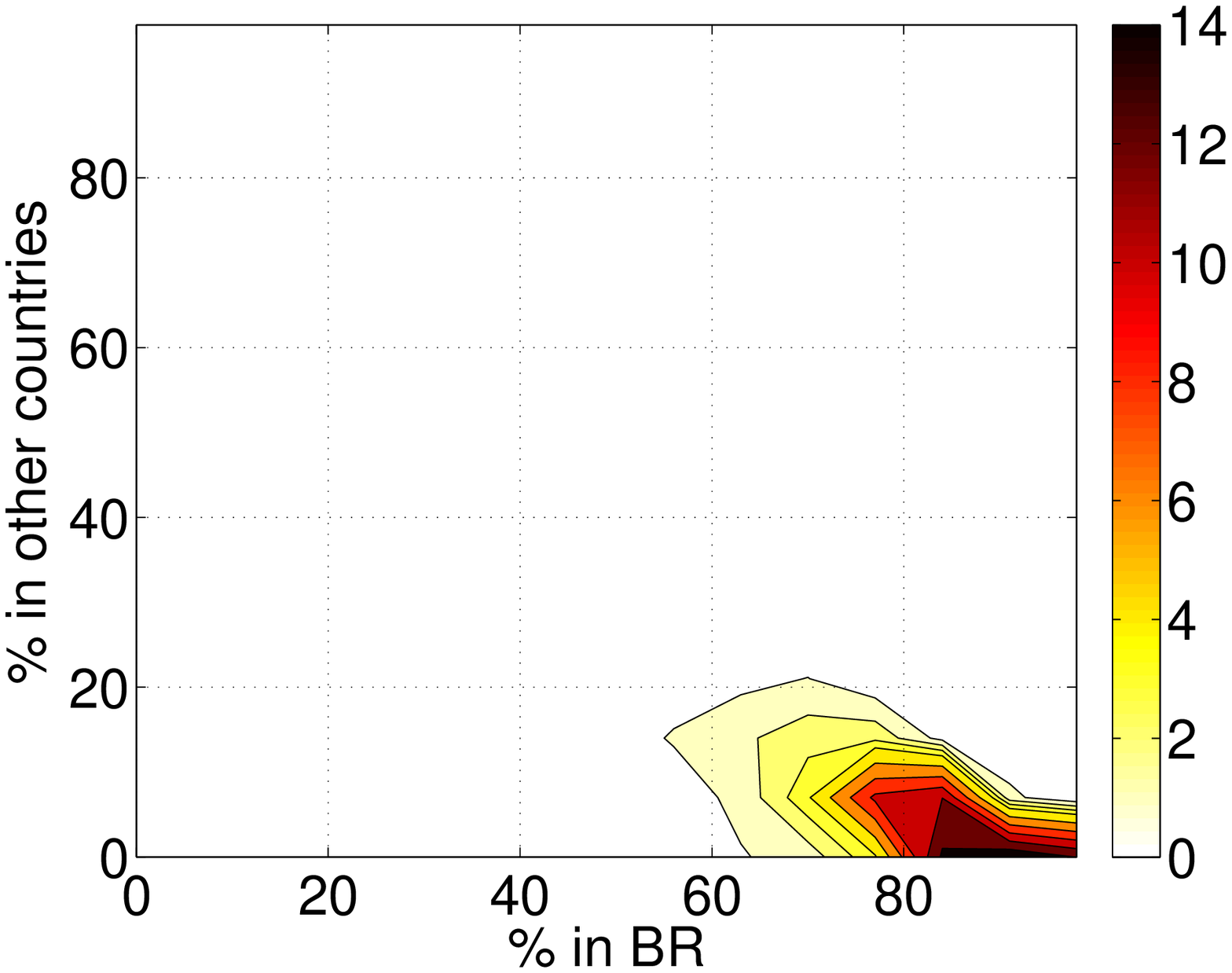} \label{fig:countries_BR_o}}

\vspace{-0.15in}
\caption{Percentage of Local followers vs Percentage of Followers in US (top) and other countries (bottom) per individual user: US, UK, France and Brazil}
\label{fig:scatter}
\vspace{-0.15in}
\end{figure*}

\section{Country Locality in Twitter}
\label{sec:country}
In this section we group the friends in our dataset by country. We have selected the country criteria since it allows to accurately group those friends having a close geographical location, a similar cultural profile and the same language. We first study the demographics of our dataset, and later perform a country-based analysis of \emph{link level} and \emph{user level} Locality.

\subsection{Twitter demographics}

In order to study the demographics of our dataset we select the 15 countries contributing a larger number of friends. The detailed demographic numbers of each one of these 15 countries are summarized in Table \ref{tab:countries}. Note that overall these 15 countries are responsible for around 90\% of our dataset. First, as already stated, we observe that US is the predominant country in Twitter responsible for around half of the friends, followers and links in our dataset. Furthermore, from the language perspective we differentiate two profiles. On the one hand, we have those countries whose official (or co-official) language is the English such as US, Canada, UK, Ireland, India and Australia. On the other hand, we find those countries with a different official language than English such as Brazil, Spain, Germany, France, Italy, Indonesia, Japan and the Netherlands. Finally, it is worth to note the presence of developing countries such as Brazil, India and Mexico in the list. This is mainly due to the high population of these countries that eases to contribute a large number of users  but also indicates the interest of their population on new social ways of communication such as Twitter.

Once we know the basic demographics of our dataset, our second aim is understanding what is the level of intra-country Locality and inter-country interaction in Twitter at link and user levels. 

\subsection{Link-based Analysis}

For each one of the Top 15 countries we compute the percentage of links originated in the country that: $(i)$ remains within the country, $(ii)$ goes to US and, $(iii)$ goes to a different country than US. Figure \ref{fig:countries_interactions} depicts the obtained results. As expected, the observed global Locality trends do not apply to every country and are mostly influenced by US Locality properties. Based on our
observations we can distinguish 4 different profiles:

\noindent \textbf{US:} due to its predominant role, it has to be considered as a separated profile. It keeps more than a 70\% of \emph{friend$\rightarrow$follower} relationships local. This is consequence of first, the predominance of US users in Twitter and second the strong local culture of US.

\noindent \textbf{Local profile}: This is formed by a group of countries keeping local a higher number of links than those going to US or other countries. This is \emph{Local} $>$ \emph{US} \& \emph{Local} $>$ \emph{Other} in Figure \ref{fig:countries_interactions}. This profile includes Brazil, The Netherlands, Indonesia, Germany and Spain. All these countries have an official language different than English. Furthermore, we found also some significant differences within the group. On the one extreme, Brazil is the country showing the highest Locality in our dataset with almost 80\% of local links. This is because it is a big country with a strong local culture and the spoken language (Portuguese) is not very spread. Just other countries, not very representative in Twitter, such as Portugal use Portuguese. On a different
corner, we have Spain whose local links are reduced  to a 41\%, since now many relations (around 30\%) are established with South-America (common language)  and other European countries (member of EU).


\noindent \textbf{Shared Locality profile}: This is formed by those countries that distribute their \emph{friend$\rightarrow$follower} links equally among those that remain local, those that go to US and those that go to other countries. This profile includes France, Mexico, Italy and Japan that are those countries where Twitter is less popular among the studied ones. Therefore, at the individual link level, intra-country Locality has a strong dependency with the local popularity of Twitter, we expect a lower intra-country Locality happening in those countries where Twitter is less popular.

\noindent \textbf{English-based (external) Locality profile}: This is formed by countries where English is the official or co-official language. These countries concentrate the major part of their links among them. Specifically, they experience an important \emph{external Locality} with many \emph{friend$\rightarrow$follower} links going to US (e.g. 48\% in the case of India and 47\% in the case of Australia and Canada). Furthermore, a lower but also important portion of links stay local (e.g. 34\% in the case of UK and 31\% in the case of Canada) and the rest are shared mainly with other English speaking countries and surrounding countries.
Therefore  in this case, we observe that the combination of language and demographics clearly influences the Locality associated to these countries. 

\subsection{User-based Analysis}

The analysis performed so far has focused on understanding the Locality at the \emph{link level}. However, as we have seen in Section \ref{sec:global} this analysis may not capture well the details at the user level. Next, we thoroughly analyze Locality at the user level for the Top 15 countries. Due to space constrains in this paper we provide the detailed analysis of one country per profile. Specifically, we consider the country with a larger number of users from each profile in our dataset. These are: US, Brazil, UK and France.

For each one of the selected countries we repeat the analysis performed in Section \ref{sec:global}. First, Figure \ref{fig:distances_countries} presents the distribution of \emph{link level} and \emph{user level} distances for each country. We confirm that in any case there is a higher Locality at the user level (curve more skewed) than at the link level. Let's now study separately each country. We observe that around 90\% of US users have typically a distance to its followers $\leq$ 4000km that defines the boundary of intra-country relationships for US. This intra-country locality effect is even more impressive in Brazil where 90\% of the users have a \emph{user level} distance $\leq$ 2000km, when the limit of intra-country relationships is also about 4000km. This confirms the presence of a regional-based Locality in Brazil.
If we analyze UK, it shows, at the user level, the bi-polarity described above between UK and US. However, contrary to the link level (34\% local, 42\% US), the user level presents a 50\% of local followers in the range of 1000Km, while those ones located in US are now reduced to a 37\%.  The second European country analyzed, France, has a 60\% of its followers closer than 1000km. However several neighbor countries such as Belgium, The Netherlands, Switzerland, Italy and Germany are located within this distance range. Hence, some portion of this 60\% represents inter-country relationships rather than intra-country ones. Finally, around 1/3 of the french users have a typical distance to its followers between 5500 and 9500 km, which represents followers population in US. Then the described shared profile is also valid at the user level.

Second, we analyze how the popularity affects the Locality for the users of each one of the studied countries. We use the same methodology explained in Section \ref{sec:global}. Figure \ref{fig:followers_countries} shows the obtained results. We observe significant differences among the countries. US shows an important correlation between popularity and Locality. The higher the popularity is the longer are the user's \emph{friend$\rightarrow$follower} links. The curves from US are similar to those observed for the whole system (See Fig \ref{fig:followers}). This is due to the preponderance of US users in Twitter, that makes the whole system showing a similar behaviour to that observed in US. Contrary, Brazil users show a high intra-country Locality (median distances around 1000km) independently of its popularity (the curve is almost flat). Finally, we can observe a clearly denoted bi-polarity in UK and France. In UK those unpopular users with less than 100 followers present a clearly marked intra-country locality, whereas the popular followers shows an \emph{external} locality phenomenon with most of its followers in the US. In France we observe the same phenomenon but the transition happens for 1000 followers.

In order to gain more insight regarding the Locality at the user level we have calculated for each individual user of these four countries the percentage of links that: stay local within the country,  goes to US and goes to a different country than US. Figure \ref{fig:scatter} depicts density diagrams in which the x-axis represents the percentage of \emph{friend$\rightarrow$follower} links that remain local and the y-axis represent the percentage of \emph{friend$\rightarrow$follower} that goes either to US (See Subfigures \ref{fig:scatter}(a,b,c)) or another country (See Subfigures \ref{fig:scatter}(d,e,f,g)) for each individual user. The results confirm and accurately quantify most of our previous observations. First we can clearly observe that the intra-country locality grows in the following way: BR $>$ US $>$ FR $>$ UK. Specifically, most of the Brazilian users have between a 80\% and 100\% of internal followers, whereas in US we observe a slightly lower intra-country locality effect since US friends present a percentage of local followers between a 70\% and 90\%. Looking at the European countries, we observe a higher level of localization in France where the vast majority of users are concentrated between 40\% and 80\% of local followers, whereas the UK shows a less concentrated diagram covering from 20\% to 80\% of local followers. Furthermore, we observe how the remote followers of UK are more concentrated in US whereas French users tend to have more followers from other countries different than US. 

\section{Related Work}
\label{sec:rw}
\noindent \textbf{Twitter Measurements:} Several previous works have exploited the different APIs offered by Twitter in order to collect data and describe different characteristic of the system. Krishnamurthy et al. \cite{bala_twitter} performed one of the initial measurement studies on Twitter collecting data of 100K users. The authors report basic characteristics of the system such as the correlation between number of followers and friends of a given user or the distribution of Twitter users per continent. Afterwards Kwak et al. \cite{sue_twitter} collected the complete \emph{friend$\rightarrow$follower} Twitter graph including 41.7 million users at the moment of the study. The authors analyze the properties of the graph topology as well as some other social aspects of Twitter such as the users influence. Also in the field of users influence Cha et al. \cite{cha_hamed_twitter} use a large dataset in order to analyze the dynamics of user influence across topic and time in Twitter. Finally, some other studies \cite{beyond_microbloging,why_we_tweet,twitter_group09} focus on understanding social aspects of the Twitter system. However, to the best of our knowledge any of the previous studies looks at neither the location of a user's followers or the Locality effect in Twitter. 

\noindent \textbf{Locality in Large Scale Applications in the Internet:} Locality is an important aspect to be considered in large scale applications. Having it into consideration may help to improve the system design and performance as has been demonstrated for the case of p2p file-sharing applications \cite{choffnes08:ONO,Cuevas2009:Deepdiving,Xie08:P4P}, p2p live-streaming applications \cite{Picconi09:locality} or OSNs such as Facebook \cite{facebook_locality}. Although Twitter has significant different characteristics than p2p applications and slightly different than Facebook, considering the Locality effect in the system design may help to improve the performance and also the data storage procedure \cite{josep_sigcomm} of Twitter.

\section{Conclusion}
\label{sec:conclusion}
Understanding the Locality effect of Internet scale systems have direct implications into the improvement and performance of such a systems. This paper is, to the best of the authors knowledge, the first study regarding the Locality phenomenon in Twitter. The obtained results demonstrate that different countries show different Locality profiles mostly influenced by the language and cultural characteristics of the country. On the one corner, we have countries with an extremely high intra-country Locality such as Brazil where most of its users keep local 80 to 90\% of the followers. On the other extreme, we have countries experiencing an external Locality phenomenon such as Australia where 50\% of the \emph{friend$\rightarrow$follower} links goes to US while just 25\% keeps local within the country. Furthermore, we have seen that US is the dominant country in Twitter responsible for around half of the friends, followers and links in our dataset. This produces that the Locality trends observed when studying the whole Twitter system are highly influenced by the Locality profile of US Twitter users.

\bibliographystyle{plain}
\bibliography{rcuevas.bib}

\appendix

\section{Accuracy of the location-tag}
\label{sec:accuracy}

In this paper we rely on the location-tag defined by the user in its Twitter profile to geolocate the user. Specifically, we are interested (for this paper) in accurately estimating the user's country. In this section, we validate the location-tag as a good approximation of the user's location. 

\begin{figure}[h]
\centering
\includegraphics[width=2.3in]{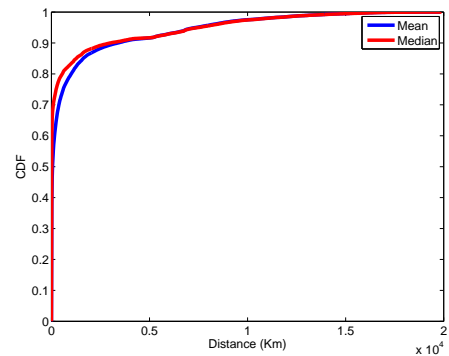} 
\caption{Median distance between the user's location-tag and the user's tweets GPS coordinates}
\label{fig:error}
\end{figure}

Twitter offers to its users the Tweet Geolocation Service. This service publishes along with the tweet the GPS coordinates from where the tweet was posted. We have collected data from 140K users that have the Tweet Geolocation Service active, have a meaningful location-tag defined in their Tweeter profile and have posted at least 5 tweets with associated GPS coordinates. For each one of these users we have computed the median geographical distance between the location specified in its Twitter profile and the GPS coordinates provided in its tweets. Figure \ref{fig:error} presents the CDF of the computed distance across the analyzed users. We can observe that most of the users ($>$ 70\%) typically post their tweets in a range of less than 100km from its specified location. Thus, we can conclude that in general the location-tag specified in the user's profile is a good estimator of the user location. Furthermore, we can consider it even more precise if we care about a correct mapping of the user to its country as we do in this paper.

\end{document}